\documentclass[sigconf]{acmart}

\usepackage{hyperref}
\usepackage{csquotes}
\usepackage{enumitem}
\usepackage{booktabs}
\usepackage{tabularx}
\usepackage{makecell}
\usepackage{tikz}
\usetikzlibrary{shapes.geometric, arrows, fit}

\setcopyright{rightsretained}
\copyrightyear{2025}
\acmYear{2025}
\acmDOI{XXXXXXX.XXXXXXX}
\acmConference[ICSE-CHASE '26]{48th International Conference on Software Engineering: Cooperative and Human Aspects of Software Engineering}{April 12--18, 2026}{Rio de Janeiro, Brazil}

\begin{document}
\title{Struggling to Connect: A Researcher’s Reflection on Networking in Software Engineering}

\author{Shalini Chakraborty}
\affiliation{%
  \institution{University of Bayreuth}
  \city{Bayreuth}
  \country{Germany}
}
\email{s.chakraborty@uni-bayreuth.de}
\orcid{0000-0002-9466-3766}

\begin{abstract}
Networking is central to the growth and visibility of software engineering research and researchers. However, opportunities and capacities to build such networks are not easily identified and often are unevenly distributed. While networking is often viewed as an individual skill, a researcher’s workplace, culture, and environment significantly influence their motivation and, consequently, the networks they form. This paper explores how factors such as country of residence, immigration status, language, gender, and surrounding context affect researchers’ ability to establish professional connections and succeed within the global research ecosystem. Drawing on existing literature and personal experience, this reflective report examines the often-invisible barriers to networking and advocates for a community-driven “expert voice” initiative to acknowledge and address these inequities.
\end{abstract}

%\begin{CCSXML}



\maketitle
\section{Introduction}
In software engineering (SE) research, success is often measured through publications, grants, and citation metrics. However, behind these indicators lies an often-overlooked enabler of research success: \textit{networking}. The professional connections that researchers cultivate can open doors to collaborations, mentorship, and recognition; all of which influence visibility and long-term impact.
Yet, not all researchers enter the field with equal opportunities to network. While digital platforms, conferences, and academic institutions aim to foster inclusivity, the reality remains that socio-geographic and personal factors deeply influence who gets to be seen and heard. 

The role of professional networks in shaping academic careers has been well-documented across disciplines. Researchers with broader networks tend to collaborate more, publish more frequently, and receive greater citation impact \cite{barabasi2002new, crane1973invisible, filipovic2023social}. In SE, collaborations often emerge through conferences, workshops, and informal professional circles \cite{storey2010impact}. These venues function not only as spaces for sharing research but also as critical social infrastructures that define who participates in the field’s intellectual discourse.

Despite this importance, networking opportunities are unequally distributed. Prior work has highlighted the structural barriers faced by early-career researchers, women, and scholars from the Global South in accessing collaborative spaces and funding networks \cite{bilecen2017introduction, mokhachane2024voices, curry2017global}. Socioeconomic and geographic disparities affect who can afford to attend international conferences, while visa restrictions and institutional funding constraints disproportionately limit researchers from low- and middle-income countries \cite{gulel2025navigating}. 
Language and cultural differences further influence participation. English remains the dominant language of publication and communication, often placing additional burdens on non-native speakers \cite{lillis2011academic}. These linguistic hierarchies shape not only who can publish but also who feels comfortable engaging in informal networking settings such as conference receptions or online discussions.

This paper presents a personal and reflective perspective on how different factors shape the capacity to connect and prosper in SE research. Rather than an empirical evaluation, this work serves as a commentary informed by literature and lived experience. Finally, it issues an open invitation for interested researchers to contribute their voices to a collaborative, community-driven report on networking inequities in academia.

\section{Inclusive Software Engineering}
Recent discussions within the SE community have called for more inclusive and transparent research practices \cite{damian2024equity,serebrenik2020diversity,hyrynsalmi2025making}. 
Damian et al.~\cite{damian2024equity} highlight how systemic inequities in participation, authorship, and leadership roles within software engineering research continue to shape who contributes to and benefits from the field’s knowledge production. 
Serebrenik et al.~\cite{serebrenik2020diversity} argue for sustained community engagement around diversity and inclusion, noting that efforts to broaden participation must be supported by transparent data reporting and accountable governance within conferences and journals. 
More recently, Hyrynsalmi et al.~\cite{hyrynsalmi2025making} have examined the cultural and structural barriers that hinder equitable collaboration, calling for inclusive research environments that recognize varied career trajectories, institutional contexts, and personal circumstances. 
These works collectively emphasize that diversity and equity are not only ethical imperatives but also key drivers of research quality and innovation. 

Despite the growing recognition of equity and inclusion as essential values in software engineering, the community still lags behind in systematically identifying and addressing the barriers that prevent equitable networking. Existing initiatives often emphasize participation metrics or demographic representation, yet less attention has been devoted to understanding the lived experiences of researchers navigating informal professional spaces \cite{damian2024equity,serebrenik2020diversity}. Prior studies have shown that academic networking remains deeply shaped by geography, language, and institutional prestige, with researchers from underrepresented regions or backgrounds facing persistent challenges in gaining visibility and collaboration opportunities. Furthermore, initiatives to build diverse and supportive networks often depend on individual agency rather than sustained structural support \cite{hyrynsalmi2025making,bjorn2022equity}. Consequently, opportunities for mentorship, collaboration, and professional growth remain unevenly distributed across the global research ecosystem. 
A major opportunity to foster inclusivity in networking lies in the spaces where professional relationships are most actively formed---academic conferences and their surrounding social ecosystems, both in-person and online.

\section{Conferences, Social Media, and the Politics of Visibility}

Academic conferences serve as the primary networking arena in SE research, but their social dynamics can reinforce existing hierarchies. Studies in computer science and related fields have shown that informal gatherings such as hallway conversations, social events, and after-hours meetups play a key role in establishing collaborations and reputations \cite{barrat2010social}. However, these interactions are often exclusionary, shaped by unspoken norms around belonging, confidence, and visibility. Access to these spaces is frequently mediated by factors such as financial resources, visa accessibility, and institutional support, which disproportionately affect early-career and underrepresented researchers \cite{gulel2025navigating, bilecen2017introduction}. 
Visa policies have become a major barrier to equitable participation in global research events. 
Current political climates and bureaucratic processes often restrict scholars from certain regions from traveling to conferences, workshops, or collaborations abroad. 
In some cases, visa processing can take more than 200 days, while conference acceptance notifications typically arrive only a few months before the event, making timely travel planning nearly impossible. 
Even when present, participants may face subtle barriers such as linguistic discomfort, social anxiety, or limited mentorship networks, which influence how effectively they can engage and be seen within the community \cite{barrat2010social}.

Most major SE conferences nowadays host informal meetups and dedicated tracks to support underrepresented participants, such as the \textit{Global South} symposium at ICSE\footnote{\url{https://conf.researchr.org/track/icse-2025/icse-2025-symposium-on-software-engineering-in-the-global-south}}. 
Additionally, co-located events such as CHASE\footnote{\url{https://conf.researchr.org/home/chase-2025}} or the New Faculty Symposium provide valuable opportunities for interaction and professional exchange. 
However, an important question remains: \emph{what happens to those who cannot attend these conferences due to financial, geographic, visa, or personal constraints? }
While online participation has grown significantly since the COVID-19 pandemic, it is questionable whether virtual formats can fully replicate the informal and serendipitous interactions of in-person conferences~\cite{guetter2022person}.
Even for those who are physically present, networking is not always straightforward. 
There are few systematic mechanisms to help newcomers or marginalized researchers connect meaningfully beyond ad hoc encounters. 
Moreover, personal traits and cognitive differences such as introversion, social anxiety, or neurodivergence can profoundly affect how participants experience large, multi-day events. 
Feeling overwhelmed in such environments is common and may inadvertently diminish one’s motivation to network or to participate in future events, particularly if the perceived outcomes do not match the personal effort and emotional energy invested \cite{sojka2024overwhelmed}

With the rise of digital scholarship, platforms such as X (formerly Twitter), Mastodon, and LinkedIn have become extensions of academic networking \cite{jordan2018academics, donelan2016social}. These platforms enable researchers to share work, build professional visibility, and engage in global discourse with fewer geographic constraints \cite{khatri2021use}. Wyrich et al.~\cite{wyrich2024beyond} investigate how software engineering practitioners and academics engage with research content on the LinkedIn platform. 
Their analysis reveals that approximately 39\% of LinkedIn posts about SE research are authored by individuals who are not the original paper authors. Interestingly, many of these posts receive little to no interaction, such as comments or discussions. This observation raises important questions about the nature of online engagement in SE, \emph{why certain research posts spark conversation while others remain unnoticed, and how these dynamics influence researchers' visibility, outreach, and perceived impact within the community?}
While social media has the potential to democratize access and flatten hierarchies, it simultaneously reproduces offline power dynamics: researchers with greater institutional prestige, linguistic fluency, or follower bases often receive disproportionate visibility and engagement \cite{veletsianos2016social, jordan2018academics}. The affordances of these platforms—such as algorithmic amplification and attention-driven metrics—tend to reward self-promotion and constant activity, privileging those with the time, confidence, and resources to maintain a sustained digital presence \cite{bozkurt2016digital, sitessocial}. Moreover, the political and emotional labor involved in managing an academic persona online can be especially taxing for marginalized scholars, who may encounter bias, harassment, or the pressure to perform identity work alongside their research contributions \cite{bozkurt2016digital, jackson2020hashtagactivism}.

Visibility in academic communities is not a neutral attribute but a socially and technologically mediated outcome. Research has long shown that recognition and influence within scientific fields are amplified through cumulative advantage—where already-visible individuals and institutions attract more attention and opportunities \cite{barabasi2002new}. In both conferences and online networks, this “Matthew effect”\cite{gomez2024matthew} manifests in who gets cited, invited to speak, or featured in community discourse. As such, visibility becomes entangled with privilege: it reflects not only scholarly merit but also the availability of resources, social capital, and structural access to key networking spaces \cite{crane1973invisible, damian2024equity}. Addressing these imbalances requires collective acknowledgment that networking is not merely a skill but a system shaped by interlocking dimensions of inequality.

\section{Factors Influencing Networking}
%\todo{How I came up with five themes? How much is personal experience vs literature?}
As members of the SE community, our experiences with networking are often mixed, ranging from rewarding collaborations to moments of exclusion or disconnect. 
However, there is a lack of studies that examine the invisible factors influencing how researchers build and sustain professional connections from their own perspectives. 

This paper adopts a reflective and interpretive methodological stance, drawing on the author’s lived experience as a software engineering researcher situated across multiple institutional, geographic and cultural contexts. Reflective scholarship has been recognized as a valid approach for surfacing tacit, under-examined aspects of research practice, particularly in areas where structural issues intersect with personal experience \cite{ellis2011autoethnography}.  

Rather than claiming generalizability, this work aims to offer analytic insight: personal experiences are used as an entry point to identify recurring patterns, which are then examined in dialogue with prior literature in software engineering, computing education and adjacent fields. The contribution lies in articulating dimensions that are often discussed implicitly, but rarely foregrounded, within SE networking discourse.

To begin addressing this gap, we propose a five-themed framework that highlights key dimensions we believe shape networking experiences in various ways. 
The five themes presented in the following section emerged through a process of reflective analysis grounded in the author’s lived experiences of networking within the software engineering research community, combined with iterative engagement with prior literature. Initial reflections focused on moments of connection, exclusion and friction encountered across conferences, collaborations and informal professional interactions. These experiences were then examined for recurring patterns and contrasted with existing empirical and conceptual work in SE and related disciplines.

\subsection{Geography and Location}
Researchers’ physical location significantly influences networking opportunities. Those based in institutions within established research hubs often benefit from proximity to influential scholars, access to well-funded research programs, and more frequent exposure to visiting academics and collaborations. Conversely, researchers situated in developing regions or remote institutions face logistical and financial barriers to attending major conferences, workshops, or informal exchanges~\cite{bozkurt2016digital}. 
This disparity, sometimes described as the \emph{``academic periphery''}~\cite{lendak2025international} effect, reinforces structural inequities that persist despite the increasing prevalence of digital communication tools. 
Even with hybrid and online participation options, researchers from the Global South or less-resourced institutions often report a sense of detachment from central academic conversations and limited visibility within the broader research ecosystem~\cite{mokhachane2024voices}.

\subsection{Immigration and Mobility}
%\todo{mention the visa issue, current political barriers for traveling}
Immigration status deeply affects both the physical mobility and social belonging of researchers. Visa restrictions, institutional sponsorship limitations, and uncertain residency status can restrict participation in international conferences, workshops, or collaborations~\cite{mokhachane2024voices}. 
These barriers disproportionately affect early-career scholars and those from low- and middle-income countries. 
Specific gendered researchers often encounter new challenges, including cultural adaptation, implicit bias, and social exclusion within host institutions~\cite{angervall2024academic}. 
Such dynamics contribute to a sense of precarity that shapes how researchers engage in professional networks and perceive their long-term academic belonging. 
Moreover, the global mobility expected in academia can inadvertently favor those with fewer family or caregiving responsibilities, thus amplifying existing inequities in who can ``move'' for opportunities~\cite{quinn2025understanding}.

\subsection{Language and Communication}
English has become the lingua franca of software engineering research, yet this linguistic dominance introduces persistent inequities. Non-native English speakers often invest additional cognitive and emotional effort in writing, presenting, and informal communication~\cite{lillis2011academic}. 
This linguistic burden can limit spontaneous participation in discussions or hallway conversations, where many collaborations and ideas emerge~\cite{al2024geopolitics}. 
Accent-based or grammar-related biases, often subtle and unintentional, can further undermine credibility or confidence during networking interactions. 
While SE is often considered a globally accessible discipline due to its shared technical vocabulary, researchers from certain regions still tend to form clusters based on linguistic familiarity, accent, and communication style. 
In their cross-cultural study involving 14 participants from ten universities across eight countries, Hoda et al.~\cite{hoda2016socio} identified language as one of the major challenges in international collaboration. 
They observed that language barriers affected not only students from non-English-speaking backgrounds but also those from native English-speaking countries, where differences in accent, idiomatic usage, and communication norms sometimes created misunderstanding or discomfort. 
Linguistic inequity in academic networking not only affects participation but also perpetuates epistemic hierarchies, where certain ways of speaking and presenting knowledge are implicitly valued over others. 
Efforts toward multilingual inclusion and translation initiatives in conferences remain limited and largely informal.

\subsection{Gender and Social Identity}
Gender continues to shape academic participation, visibility, and access to professional networks. 
Studies have shown that women in computing and software engineering experience greater barriers to recognition, mentorship, and inclusion in informal professional circles~\cite{vasilescu2015gender}. 
Such barriers are often perpetuated through microaggressions, homophily in collaborations, and exclusion from informal networking spaces such as conference dinners or social events~\cite{ford2016paradise}. 
Intersectional factors such as race, ethnicity, disability, and sexual orientation, compound these inequities, leading to unique challenges for those navigating multiple marginalized identities~\cite{damian2024equity,serebrenik2020diversity}. 
In a recent study Boman et al.~\cite{boman2024breaking} discuss social barriers such as imposter syndrome, tokenism,  conditional belonging, etc for women and non-binary students in SE. 
Even when formal inclusion initiatives exist, implicit biases and unequal labor expectations (e.g., the “service burden” often carried by women and minoritized academics) can undermine their intended effects~\cite{damian2024equity,hyrynsalmi2025making}. 
A holistic understanding of networking inequities must therefore account for these overlapping social dimensions and their systemic roots in academic culture.

\subsection{Personality, Disposition, and Neurodiversity}
Networking practices often privilege extroverted, socially assertive, or self-promotional behaviors, traits that align with Western academic norms of confidence and visibility. 
Introverted researchers may find traditional networking environments, such as receptions or poster sessions, uncomfortable or overstimulating~\cite{cain2016quiet}. 
Similarly, individuals with social anxiety or differing cognitive styles can experience significant emotional fatigue during multi-day events that demand constant interaction and impression management~\cite{bryant2025diverse}. 
For neurodivergent scholars (e.g., those on the autism spectrum, with ADHD, or other cognitive differences), academic conferences can pose additional sensory, communicative, or organizational challenges that affect participation~\cite{rollnik2024managing}. 
While online and asynchronous formats may offer more accessible alternatives for some, they also introduce new complexities around self-presentation, visibility, and engagement expectations. 
Recognizing neurodiversity as a component of inclusion highlights the need for more flexible, multi-modal approaches to networking—ones that value diverse cognitive and social dispositions rather than rewarding a single model of engagement.

\section{The Struggle of Researchers}
The interplay of the above factors creates a complex and uneven landscape of opportunities within the software engineering research community. 
% NEW
In SE, geographic constraints are particularly consequential due to the field’s strong reliance on conference-centered networking, collaborative tool-building and industry-facing research. Unlike some disciplines where publication alone may suffice for visibility, SE often privileges sustained interpersonal interaction—through hallway conversations, artifact demonstrations, workshops and repeated community presence—which amplifies the impact of location-based barriers.
Researchers from underrepresented regions or identities may find themselves unintentionally excluded from influential collaborations or unaware of informal channels that shape career trajectories. 
Limited access to mentorship and professional networks further amplifies these disparities, reinforcing structural inequities that persist over time. 
Even researchers affiliated with well-known institutions may face challenges when cognitive or personality differences hinder their ability to engage effectively in traditional networking spaces.

Emotional labor adds another layer to this complexity. 
Constantly navigating unfamiliar cultural, linguistic, or social environments can lead to fatigue, imposter syndrome, and feelings of isolation. 
For early-career researchers, particularly those transitioning to new countries or institutions, the effort to ``fit in'' within established communities can sometimes overshadow the joy of discovery and collaboration that defines academic life.
Despite growing awareness of equity, diversity, and inclusion, academic structures continue to privilege those already embedded within established networks of visibility and influence. 
Roles such as program committee membership, and conference organization not only strengthen one’s professional standing but also provide learning and leadership experiences that build confidence and belonging. 
However, access to these opportunities remains heavily dependent on prior connections, perpetuating a cycle where networking itself becomes both the means and the gatekeeper of success.

\section{Call for a Community Effort}
%\todo{Exactly what are the takeaways: how to implement the study and get evidence (future work)}
Software engineering, by its very nature, offers a unique opportunity to connect people across linguistic, cultural and geographic boundaries. The discipline’s reliance on shared technical languages from programming syntax to design diagrams creates a form of communication that can transcend spoken language barriers and foster mutual understanding through practice. 
Moreover, the field’s strong ties between academia and industry, coupled with the increasing prevalence of remote and distributed work, make SE an inherently collaborative and globally accessible domain.

Addressing networking inequities in software engineering requires collective, community-level action rather than reliance on individual coping strategies. This paper is not intended as a conclusion but as a starting point. The next step is to bring together a community of researchers interested in co-authoring an ``expert voice'' report that systematically documents the experiences, challenges and coping strategies of researchers across diverse backgrounds regarding networking.

Based on the themes discussed, we outline preliminary directions for action that are particularly relevant to SE:

%The envisioned report will:
%\begin{itemize}
%    \item Identify and analyze the structural and personal barriers of networking in SE research.
%    \item Highlight stories of resilience and strategies that enable success despite such barriers.
%    \item Provide recommendations for conferences, institutions, and funding bodies to foster inclusive networking environments.
%\end{itemize}
\begin{itemize}
    \item \textbf{Make networking structures visible:} SE venues can explicitly document how collaborations, program committee roles, workshops and informal influence networks typically form, reducing reliance on tacit knowledge.
    \item \textbf{Design inclusive interaction formats:} Conferences and workshops can complement informal social events with structured, low-barrier networking opportunities that accommodate diverse communication styles. Existing initiatives such as \emph{new faculty symposium} and \emph{mentoring workshops} already help bring participants together; however, additional formats—such as open problem discussion sessions or remote and hybrid mentoring programs that support participation across geographic locations—are needed.
    \item \textbf{Value diverse forms of participation:} SE communities can broaden recognition beyond visibility-heavy roles (e.g., frequent speakers) to include behind-the-scenes contributions such as artifact development, mentoring, and community maintenance.
    \item \textbf{Support evidence-building efforts:} The community should encourage empirical and mixed-methods studies that systematically examine networking experiences in SE, using reflective accounts such as this work as a starting point for hypothesis generation.
\end{itemize}

% NEW
By situating networking as an infrastructural concern rather than an individual shortcoming, the SE community can begin to align its collaborative practices with its stated commitments to equity, inclusion and sustainability.
Interested researchers from all backgrounds, especially those from underrepresented or geographically distant regions, are warmly invited to participate. Together, we can acknowledge these struggles, amplify overlooked voices and inspire a more equitable and connected software engineering community.

\newpage
\bibliographystyle{acm}
\bibliography{refs}
\end{document}